\documentclass[a4paper]{jpconf}
\usepackage{graphicx}    
\usepackage{amsmath}
\usepackage{esint}
\usepackage{array}       
\usepackage{verbatim}    
\usepackage{color}
\usepackage{colortbl}
\usepackage{listings}
\usepackage{epstopdf}
\usepackage{subfigure}
\usepackage{rotating}
\usepackage[percent]{overpic}
\usepackage{wrapfig}
\usepackage{hyperref}
\hypersetup{pdfborder={0 0 0}, colorlinks=true, citecolor=blue, linkcolor=blue}
\usepackage{url}
\usepackage{tikz}  
\usetikzlibrary{shapes,arrows,positioning,backgrounds,calc}
\tikzstyle{flow} = [rectangle, rounded corners, text centered, draw=black, fill=white, ultra thick,align=left]

\definecolor{darkred}{RGB}{175,0,0}

\begin{document}
\title{GPU Acceleration of an Established Solar MHD Code using OpenACC}
\author{R. M. Caplan, J. A. Linker, Z. Miki{\'c}, C. Downs, T. T{\"o}r{\"o}k, and V. S. Titov}
\address{Predictive Science Inc., 9990 Mesa Rim Road Suite 170, San Diego, CA  92121}
\ead{caplanr@predsci.com, linkerj@predsci.com, mikicz@predsci.com, cdowns@predsci.com, ttorok@predsci.com, titovv@predsci.com}

\begin{abstract}
GPU accelerators have had a notable impact on high-performance computing across many disciplines.  They provide high performance with low cost/power, and therefore have become a primary compute resource on many of the largest supercomputers.  Here, we implement multi-GPU acceleration into our Solar MHD code (MAS) using OpenACC in a fully portable, single-source manner.   Our preliminary implementation is focused on MAS running in a reduced physics ``zero-beta" mode.  While valuable on its own, our main goal is to pave the way for a full physics, thermodynamic MHD implementation.  We describe the OpenACC implementation methodology and challenges.  ``Time-to-solution" performance results of a production-level flux rope eruption simulation on multi-CPU and multi-GPU systems are shown.  We find that the GPU-accelerated MAS code has the ability to run ``zero-beta" simulations on a single multi-GPU server at speeds previously requiring multiple CPU server-nodes of a supercomputer.
\end{abstract}

\section{Introduction}
\label{sec:intro}

Hardware accelerators (e.g.~graphics processing units (GPUs)) have significantly impacted the state-of-the-art of high-performance computing (HPC) in the fields of AI and machine learning, as well as in traditional scientific applications \cite{GPUNATURE2018}.  GPU-accelerated codes can achieve performance normally only available on super computers using a single `in-house' system with a modest number of accelerator cards (often at much lower cost).  GPUs exhibit drastic improvements in power efficiency {--} a key reason why many of the largest HPC systems in the world are based around them.  In some cases, they can provide necessary performance that cannot be achieved with traditional CPU systems.  

Originally, the only way for software to take advantage of GPU accelerators was to write/rewrite code using APIs such as CUDA \cite{cook2012cuda} and OpenCL \cite{kaeli2015heterogeneous}.  However, for legacy scientific applications, rewriting code can be far too difficult in terms of development time, effort, and education of the user base.  Additionally, because portability and longevity is of critical concern, the wide-scale adoption of accelerated programming into such applications has been relatively slow.

Programming for accelerators through the use of directives addresses the above concerns.  Accelerator directives are expressed as special comments in the code.  This allows a single source code to be compiled for both accelerators (using a directive-supporting compiler), and CPUs (using any CPU compiler).  For NVIDIA GPUs, the most common directive-based API in production is the open standard OpenACC\footnote{\url{http://www.openacc.org}} \cite{openaccbook1,openaccbook2}. Implementations of OpenACC can also compile to other device types \cite{openaccbook2} including FPGAs, multi-core CPUs, Power9, and SunWay processors \cite{fu2016sunway}.  The most complete implementation of OpenACC is currently in the PGI compiler from the Portland Group\footnote{\url{http://www.pgroup.com}}. There are also open-source OpenACC implementations, including the GNU compiler\footnote{\url{https://gcc.gnu.org/wiki/OpenACC}}, but these currently lack some features. In this paper, we will use PGI's OpenACC implementation to accelerate our code for use on NVIDIA GPUs.  We note that OpenMP\footnote{\url{http://www.openmp.org}} has also recently added accelerator support into its API, but we opted for the maturity and relative simplicity of OpenACC for this work.

\begin{figure}[htbp]
\centering
\begin{tikzpicture}[node distance=0.5in]
\node (orig) [flow] { 
$\begin{array}{c}\mbox{Original Code}\\\hline\end{array}$\\
\begin{lstlisting}
for (i=0; i<N; i++)
   y[i] = a*x[i] + y[i];
\end{lstlisting}
};
\node (acc) [flow, below of=orig,node distance=1.5in] {
$\begin{array}{c}\mbox{{\color{cyan}OpenACC} GPU Code}\\\hline\end{array}$\\
\begin{lstlisting}
#pragma acc enter data copyin(x,y)
{
#pragma acc kernels present(x,y)
for (i=0; i<N; i++)
   y[i] = a*x[i] + y[i];
}
#pragma acc exit data copyout(y)   
\end{lstlisting}
};
\node (cuda) [flow, right=of orig, anchor=north west] {
$\begin{array}{c}\mbox{{\color{green}CUDA} GPU Code}\\\hline\end{array}$\\
\begin{lstlisting}
__global__ void saxpy(int N,float a,
                      float* restrict x,
                      float* restrict y){
  int i=blockIdx.x*blockDim.x+threadIdx.x;
  if (i < N) y[i]=a*x[i]+y[i];
}
...
const int BLOCK_SIZE=2048;
float *d_x,*d_y;
...
cudaMalloc( (void **) &d_x, sizeof(float)*N);
cudaMalloc( (void **) &d_y, sizeof(float)*N);
cudaMemcpy(d_x, x, N, cudaMemcpyHostToDevice);
cudaMemcpy(d_y, y, N, cudaMemcpyHostToDevice);

dim3 dimBlock(BLOCK_SIZE);
dim3 dimGrid((int)ceil((N+0.0)/dimBlock.x));

saxpy<<<dimGrid,dimBlock>>>(N,a,d_x,d_y);

cudaMemcpy(y,d_y,N,cudaMemcpyDeviceToHost);
cudaFree(d_x);
cudaFree(d_y);
\end{lstlisting}
};
\draw [-stealth', ultra thick] (orig) -- (cuda);
\draw [-stealth', ultra thick] (orig) -- (acc);
\end{tikzpicture}
\caption{Example GPU implementations of a SAXPY algorithm using CUDA and OpenACC.  The OpenACC implementation is simpler and maintains portibility.\label{fig:code}} 
\end{figure}
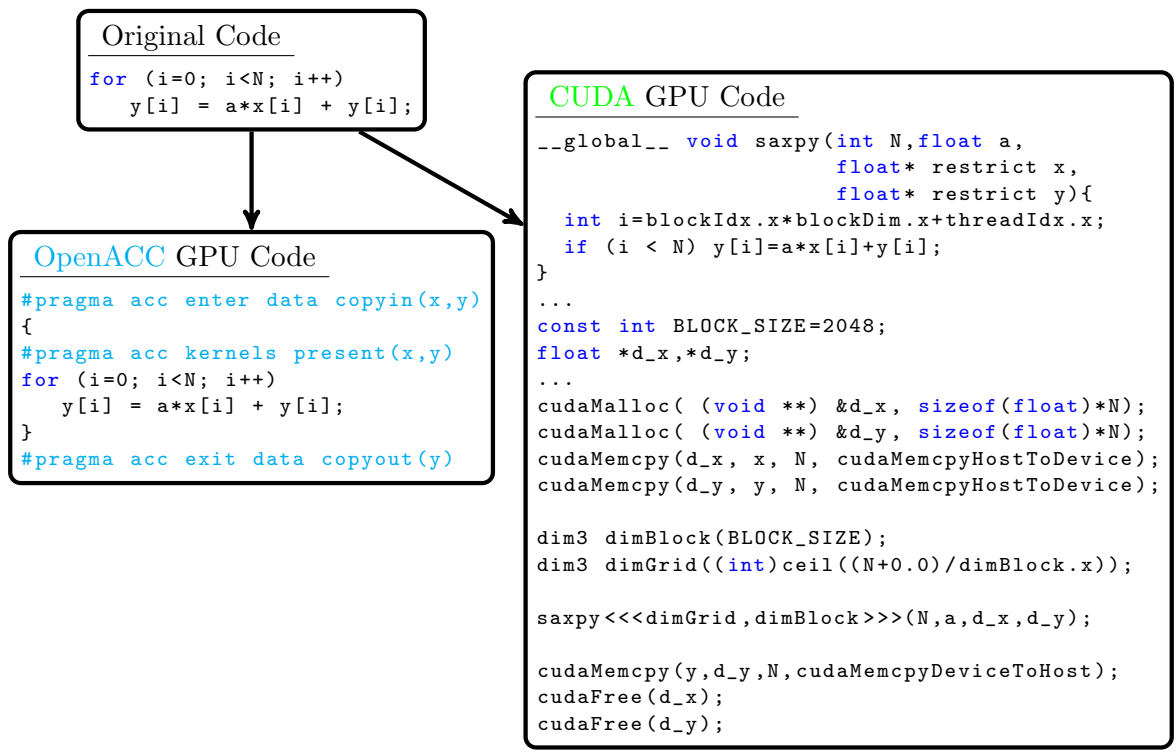

In Figure.~\ref{fig:code}, we illustrate the coding advantages of OpenACC over traditional GPU language extensions, by converting a simple SAXPY code segment for use on NVIDIA GPUs using both CUDA and OpenACC.
We see that the OpenACC implementation requires much less additional code, preserves the original code, and is backwards compatible with the original code.

In this paper, we show our initial implementation of OpenACC into our Solar Magnetohydrodynamic (MHD) code MAS (described in Sec.~\ref{sec:mas}), allowing it to run on multiple NVIDIA GPUs.  Our goal in porting MAS to GPUs is to allow for multiple small-to-medium sized simulations to be run on an in-house multi-GPU server, greatly benefiting development and parameter studies. In addition, very large simulations using MAS take up a substantial portion of our HPC allocations. Having the code run on multiple GPU nodes can decrease the time-to-solution using the same number of nodes and/or reduce the total allocation usage.

The paper is organized as follows: in Sec.~\ref{sec:mas} we introduce the MAS code and describe its use, models, and algorithms.  We then describe the production-level test simulation we use for timing results in Sec.~\ref{sec:testcase}. The OpenACC implementation, including algorithm analysis, difficulties encountered, and an overall summary of the effort, is discussed in Sec.~\ref{sec:imp}.  In Sec.~\ref{sec:timings} we show extensive timing results across multiple algorithms and hardware models (both GPUs and CPUs), and we conclude in Sec.~\ref{sec:conclusion}.

\section{The MAS Code}
\label{sec:mas}
Magnetohydrodynamic Algorithm outside a Sphere (MAS) is an in-production MHD code with over 15 years of ongoing development used extensively in Solar physics research \cite{mikic99a,linker99a,downsetal2013,lionello05a,lionello06a,linkeretal2011,linker03a,lionelloetal2013}.  It was used to predict the coronal structure during the August 21st, 2017 total solar eclipse \cite{ECLIPSE2017} and has demonstrated the ability to reproduce qualitative and quantitative properties of solar eruptions \cite{torok18}.  The code is currently included as a part of the CORHEL suite \cite{rileyetal2012} hosted at the Community Coordinated Modeling Center (CCMC)\footnote{\url{http://ccmc.gsfc.nasa.gov}} which allows users to generate background MHD solutions of the corona and heliosphere based on observational boundary conditions. MAS is written in FORTRAN ($\sim 50,000$ lines) and parallelized with MPI.  It can run simulations containing over two hundred million grid cells and exhibits performance scaling to thousands of CPU cores \cite{ASTRONUM16}.  

\subsection{Models}
MAS solves a well-featured thermodynamic resistive MHD model with numerous physics and algorithm options \cite{MAS,ASTRONUM16}.  However, not all features and model equations are used for every simulation. The so-called `zero-beta' mode freezes the density and sets the pressure to zero, essentially eliminating several evolution equations.  Zero-beta simulations are applicable in low-beta regions of the corona (i.e. where the magnetic forces dominate the dynamics) and can have many uses including the generation of initial solutions for solar storm magnetic structures and analyzing initial eruption dynamics \cite{torok18}.  The zero-beta model is described as
\begin{equation}
\begin{aligned}
\frac{\partial {\bf A}}{\partial t} &={\bf v}\times \left(\nabla\times {\bf A}\right)-{\frac{c^2\,\eta}{4\,\pi}\,\nabla \times \nabla \times {\bf A}} 
\\
\frac{\partial{\bf v}}{\partial t}& = -{\bf v}\cdot \nabla\,{\bf v} + \frac{1}{\rho}\left[ \frac{1}{c}{\bf J} \times {\bf B}\right] + {\frac{1}{\rho}\,\nabla\cdot(\nu \rho \nabla {\bf v})} + {\frac{1}{\rho}\,\nabla\cdot\left(S\,\rho\,\nabla{\frac{\partial \bf v}{\partial t}}\right)} 
\\
p&=0 
\\
\rho &= \rho_0({\bf r}) 
\end{aligned}
\end{equation}
where ${\bf A}$ is the magnetic vector potential, ${\bf B}=\nabla\times {\bf A}$ is the magnetic field, ${\bf J}=\frac{c}{4\,\pi}\nabla\times {\bf B}$ is the current density, $\rho$ is the plasma density, $p$ is the plasma pressure, and {\bf v} is the plasma velocity, $\eta$ is the coefficient of resistivity, $\nu$ is the coefficient of kinematic viscosity, and $S$ is a semi-implicit factor defined as $S=(\Delta t^2\,\tilde k^2)^{-1}\,(C_w^2/(1-C_f)^2-1)$, where $\tilde k^2$ is the upper limit of the combined inverse grid spacing, $C_f$ is the flow CFL, and $C_w$ is the fast magnetosonic wave CFL \cite{ASTRONUM16}.   The resistivity and viscosity operators are added in order to avoid the potential emergence of oscillatory behavior due to the algorithms in MAS not guaranteeing monotonicity.  The semi-implicit operator is added in order to be able to advance the code at the flow CFL time-step stability limit which is much larger than the fast magneto-sonic wave limit (see Ref.~\cite{ROBSTB} for details).

Since the computations involved in running the zero-beta model contain all the major algorithms and data structures used in the fully-featured mode of MAS, the zero-beta model makes an ideal first step for our OpenACC implementation of MAS and is the focus of this paper.

\subsection{Algorithms}
\label{sec:algs}
MAS solves the MHD equations on a non-uniform logically-rectangular staggered grid using finite difference algorithms.   Several time integration schemes are used for various parts of the MHD model.  Advective and reactive terms are integrated with an explicit predictor-corrector upwinding scheme \cite{ROBSTB}, the semi-implicit term is solved in the predictor-corrector framework implicitly using two preconditioned conjugate gradient (PCG) solves, while diffusive-like terms are either solved implicitly with PCG solves or explicitly with a Runge-Kutta Legendre super time-stepping (STS) scheme \cite{RKL2_2014}.  
The PCG solver in MAS has two communication-free preconditioning options.  The first (PC1) uses the inverse diagonal of the matrix (point-Jacobi), while the second (PC2) is a block-Jacobi using a zero-fill incomplete LU factorization \cite{IterativeMethods_SAAD_Book}.  For `easy' solves (low condition number), PC1 typically runs faster than PC2, while for `hard' solves (high condition number) PC2 can be much faster than PC1.  The resistivity solve is typically always `easy' and therefore always uses PC1.  For the semi-implicit and viscosity solves in coronal simulations, PC2 is usually faster than PC1, but PC1 scales more efficeintly to high numbers of processors than PC2.  The STS algorithm for viscosity typically runs as fast (often faster) than PCG+PC2, and scales better than PCG even with PC1.  However, it can suffer from accuracy-breakdown in certain problem configurations (see Ref.~\cite{ASTRONUM16} for details).   
For the implementation and testing in this paper, we define the following four algorithmic choices:
\begin{itemize}
\item[1)] Semi-Implicit: PCG+PC1, Viscosity: PCG+PC1
\item[2)] Semi-Implicit: PCG+PC2, Viscosity: PCG+PC2
\item[3)] Semi-Implicit: PCG+PC1, Viscosity: STS
\item[4)] Semi-Implicit: PCG+PC2, Viscosity: STS
\end{itemize}

\section{Real-world Test Case}
\label{sec:testcase}
\begin{figure}[b]
\centering
\includegraphics[height=1.50in]{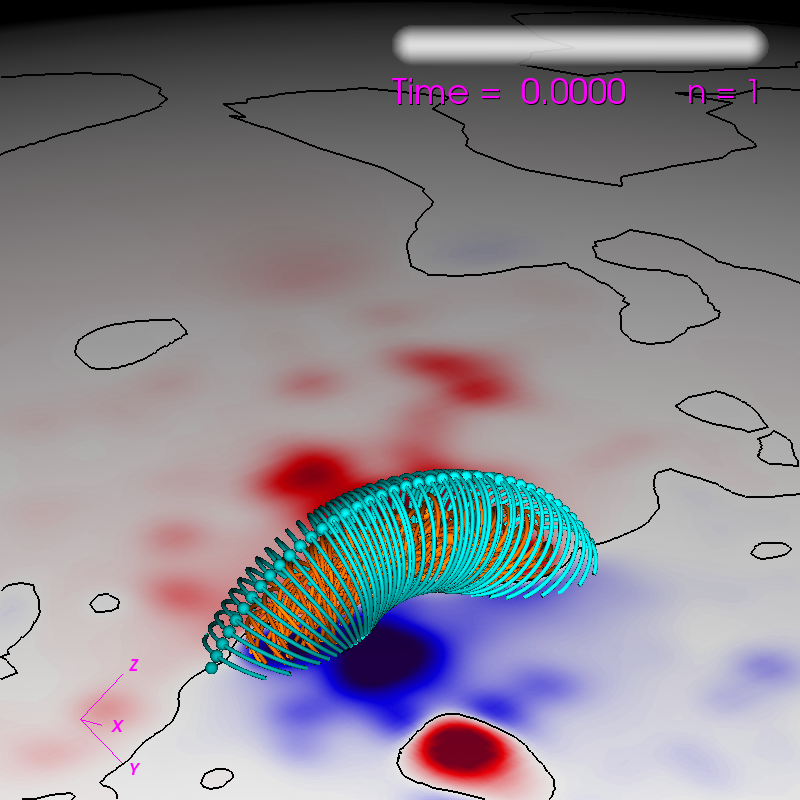}
\includegraphics[height=1.50in]{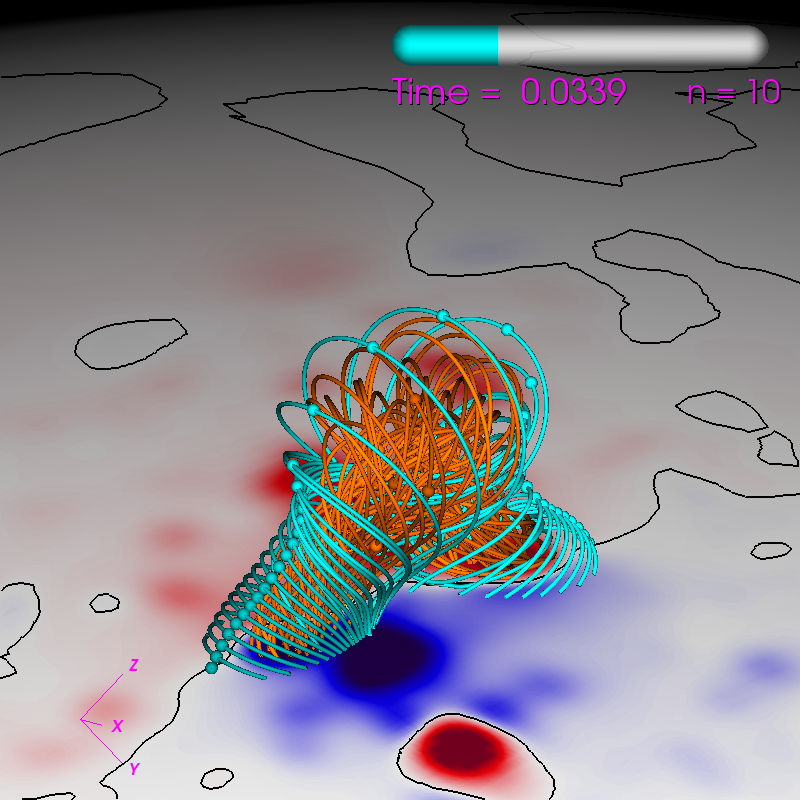}
\includegraphics[height=1.50in]{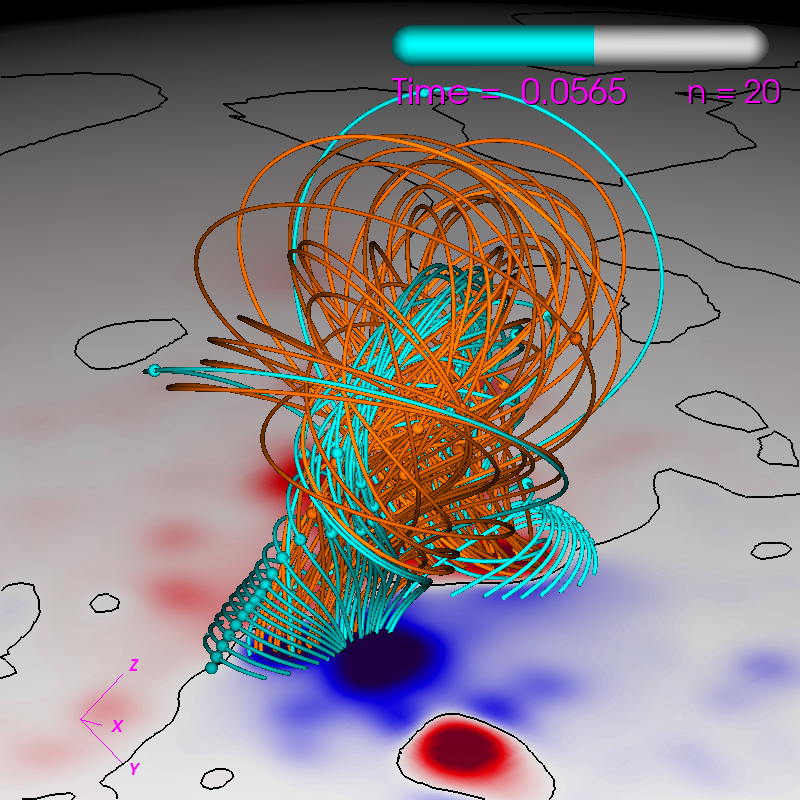}
\includegraphics[height=1.50in]{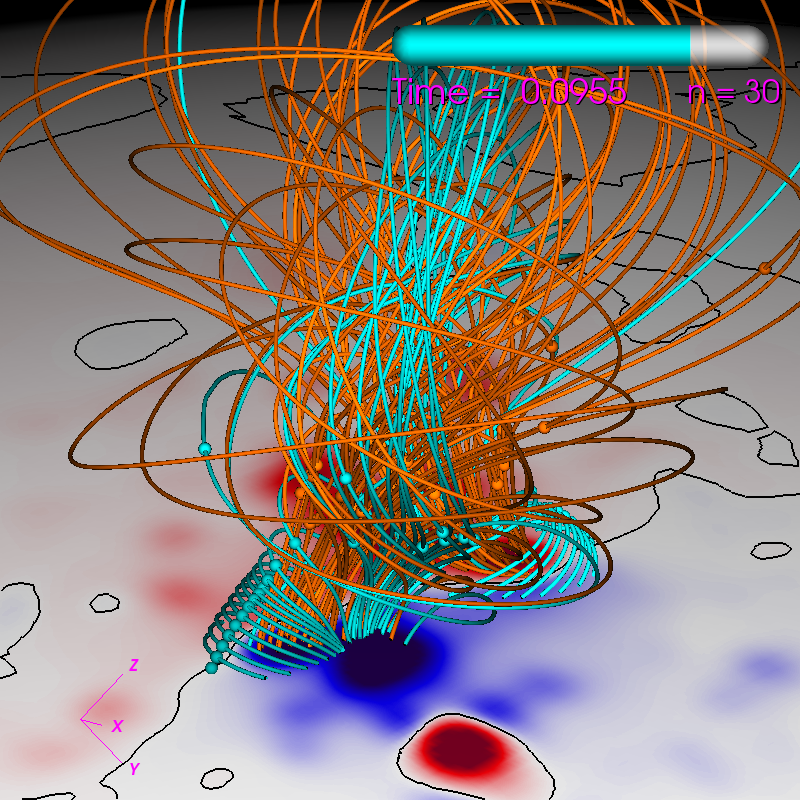}
\caption{Magnetic field evolution from the simulation used to test code performance.  The orange and cyan lines represent magnetic field lines, while the colormap represents the radial magnetic field values on the solar surface.\label{fig:testcase}} 
\end{figure}

In order to provide useful timings, we select a full production run of MAS. We use a global zero-beta simulation of the onset of a solar eruption with a TDm flux rope \cite{titov2014method} structure (see Ref.~\cite{mikic2013challenge} for similar simulations). The domain extends all around the Sun (avoiding the need to impose artificial boundary conditions) from the solar surface out to 10 solar radii. It uses an extremely non-uniform grid of $\sim$11 million cells, with very small cell sizes in the erupting region, coarsening greatly with distance. We therefore employ a large, uniform, viscosity coefficient sufficient to ensure stability in the very coarse regions. The simulation covers a physical evolution of $\sim$200 seconds requiring $\sim$700 non-uniform time steps. Figure~\ref{fig:testcase} shows visualizations of the run.
The semi-implicit PCG solves take an average of $\sim$70 iterations per time step with the PC1 preconditioner and between $\sim$30{--}50 iterations using PC2 (the PC2 preconditioner degrades in effectiveness as the number of MPI ranks increases).  The viscosity solve is more difficult (due to the large viscosity coefficient), requiring an average of $\sim$550 iterations using PCG+PC1, and between $\sim$100{--}200 iterations using PCG+PC2.  Advancing viscosity with the explicit STS algorithm requires an average of $\sim$200 sub-steps per time-step.   

\subsection{Validation}
\label{sec:valid}
\begin{wrapfigure}{r}{0.5\textwidth}
\centering
\includegraphics[width=0.47\textwidth]{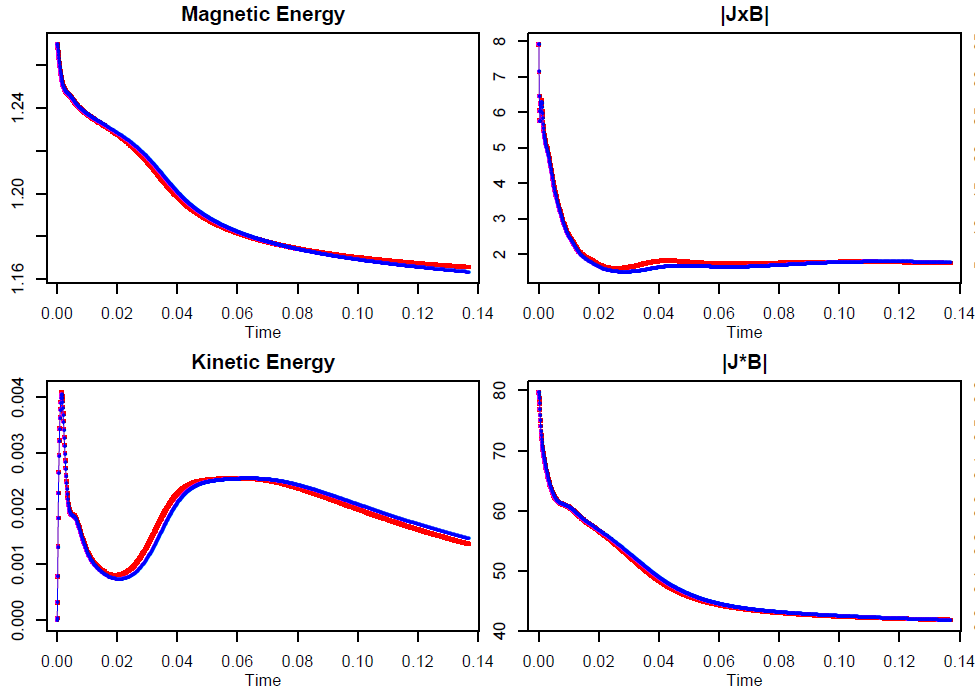}
\includegraphics[width=0.16\textwidth]{figures/fl_iso020.png}
\includegraphics[width=0.16\textwidth]{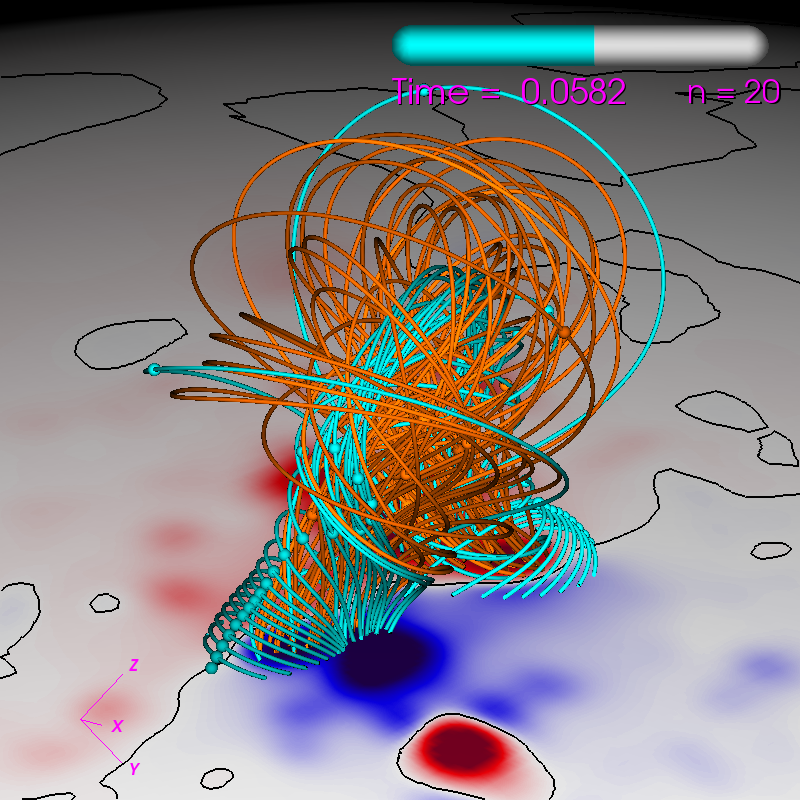}
\caption{Comparison of the test simulation using PCG+PC2 and STS for viscosity.  Top two rows: Volume integrated quantities (in code units) between PCG+PC2 (red) and STS (blue). Bottom row: Magnetic field visualization of the run with viscosity solved by PCG+PC2 (left) and STS (right).\label{fig:validation}} 
\end{wrapfigure}
As mentioned in Sec.~\ref{sec:algs}, using the STS algorithm for the viscosity operator is often faster than using PCG+PC2.  In Ref.~\cite{ASTRONUM16}, we showed that the STS algorithm was over twice as fast as PC2+PCG in a thermodynamic MHD coronal relaxation simulation.  However, in that case, the STS algorithm exhibited a break-down in accuracy, causing nonphysical oscillations in high resolution areas.  Therefore, we must carefully validate the solution when deciding whether or not to use the STS method.  In Figure.~\ref{fig:validation} we show values of the total magnetic and kinetic energies, and the total $|J\times B|$ and $|J\cdot B|$ for the test simulation of Sec.~\ref{sec:testcase} using both the PCG+PC2 and STS algorithms, along with a side-by-side comparison of visualizations.  We see that the two simulations yield very similar results and the solutions appear nearly indistinguishable.  Therefore, for this example problem, we can safely include STS as a valid choice of algorithm for viscosity.  

Our GPU runs described below were also validated with the original CPU runs.  The GPU runs matched the CPU runs quantitatively to within the PCG solver tolerance (due to floating-point variations).

\section{Implementing OpenACC in MAS}
\label{sec:imp}

Here we summarize our implementation of OpenACC into MAS.  Further details, including specific code examples, strategies and development tips, can be found in Ref.~\cite{CAPLAN_DEVBLOG}.

\subsection{Algorithm Analysis}
\label{sec:alganalysis}
GPUs are very efficient for a variety of algorithms, but not all.  Therefore it is important to analyze the algorithms and determine if they are suitable for GPU acceleration. It is key that an algorithm is vectorizable (i.e. that it can perform a single operation on a contiguous section of an array in memory (SIMD)).  Basic array operations (e.g. AXPY) are vectorizable, as are explicit finite difference algorithms (such as the stencil operations in the advection operators and the STS algorithm). The core operators in our PCG solvers consist of sparse matrix-vector products (essentially stencil operations), basic array operations, and dot products, all of which are vectorizable to varying degree. The PC1 preconditioner consists of a simple array operation, but the PC2 preconditioner relies on a local triangular matrix solve computed using a standard backward-forward sequential algorithm which is not vectorizable. To address this issue, NVIDIA has developed the CUDA-based cuSparse library containing advanced (and vectorizable) triangular sparse solvers for GPUs.  Although OpenACC codes are able to link to CUDA libraries (like {\tt cuSparse}), doing so breaks portability as the libraries often do not contain CPU versions of the algorithms.  To test if this loss of portability is outweighed by the performance advantage of PC2, we implemented {\tt cuSparse} into our OpenACC GPU-accelerated potential field solver POT3D \cite{POT3DACC}.  The results showed that solving our symmetric matrices (which have few off-diagonal non-zero entries) on modern GPUs with PC2 yielded similar performance as using PC1.  Therefore, we choose to prioritize portability and do not GPU-accelerate the PCG+PC2 algorithm in MAS (limiting the algorithm choices in Sec.~\ref{sec:algs} to (1) and (3) for use on GPUs).

\subsection{Examples of OpenACC}
\label{sec:accexamples}
\lstset{language=Fortran,
  breaklines=true,
  basicstyle=\fontsize{8}{10}\ttfamily,
  keywordstyle=\color{black},
  commentstyle=\color{cyan},
  frame = single
}
A full description of the OpenACC directives used to port the MAS code is beyond the scope of this paper.  We refer the reader to the many available books and tutorials on OpenACC programming \cite{openaccbook1,openaccbook2}.  Here, we describe some of the more fundamental concepts through selected examples.

\begin{wrapfigure}{r}{0.5\textwidth}
\centering
\noindent\begin{minipage}{.475\textwidth}
\begin{lstlisting}
     allocate/initialize "y" on CPU ...
!$acc enter data copyin (y)
     GPU version of "y" now available for     
     use in OpenACC compute regions ...
!$acc update self (y)
     CPU version of "y" updated 
     (for I/O, etc.) ...
!$acc exit data delete(y)
\end{lstlisting}
\end{minipage}
\caption{Outline for using unstructured data regions in OpenACC.\label{fig:accexp1}} 
\end{wrapfigure}
One key aspect of efficient GPU programming is carefully managing data movement.  GPUs have their own local memory and data needs to be first transferred to the GPU for computation, and then retrieved if needed by the CPU.  This transfer has very high latency and needs to be avoided as much as possible.  For MAS, we use OpenACC's unstructured data regions to manually send the data to the GPU, only transferring it back to the CPU for I/O.  This is shown schematically in Figure.~\ref{fig:accexp1}.  
\lstset{language=Fortran,
  breaklines=true,
  basicstyle=\fontsize{8}{10}\ttfamily,
  keywordstyle=\color{blue},
  commentstyle=\color{cyan},
  frame = single
}

In order to avoid data transfers except in the case of I/O, we needed to apply OpenACC directives to every part of the code that uses the data, resulting in a large number of added directives.  In Figure~\ref{fig:accexp2}, we show examples of adding OpenACC directives to two basic code loops - an AXPY operator and a summation operator.  
\begin{figure}[htbp]
\centering
\begin{tabular}{cc}
\noindent\begin{minipage}{.4\textwidth}
\begin{lstlisting}
!$acc kernels loop default(present)
   do i=1,n
     y(i) = a*x(i) + y(i) 
   enddo
!$acc end kernels
\end{lstlisting}
\end{minipage}
&
\noindent\begin{minipage}{.55\textwidth}
\begin{lstlisting}
!$acc kernels loop present(y) reduction(+:total) 
   do j=1,m
     total = total + y(j)
   enddo
!$acc end kernels   
\end{lstlisting}
\end{minipage}
\end{tabular}
\caption{Basic examples of applying OpenACC to an AXPY operator loop (left) and a summation (right).\label{fig:accexp2}} 
\end{figure}
The {\tt default(present)} and {\tt present()} directives indicate that the required data has already been  transferred to the GPU and are available.  In the summation loop, {\tt total} is needed to be declared as a reduction variable along with the reduction operator (in this case ``+").  

For large problems, it is critical to be able to compute on multiple GPUs - both within a server node and between nodes on a cluster.  The MAS code splits the domain of the simulation and gives each MPI rank (typically assigned to one CPU core) a section of the domain to work on.  When one rank requires data from other ranks, MPI communications are used to send and receive the required data.  To use multiple GPUs, we assign MPI ranks so that there is one CPU rank per GPU.  We then use OpenACC's {\tt set device\_num} directive to tell the run-time to use the desired GPU.  For this to work across a cluster, the code needs to know the local MPI ranks on a given node.  To do this we use MPI-3's shared communicator feature as shown in Figure~\ref{fig:accexp3}.  Figure~\ref{fig:accexp3} also shows an example of how we avoid CPU-GPU transfers when using MPI communication by using the {\tt host\_data use\_device} clause in OpenACC to call MPI routines with GPU data directly (this requires a `CUDA-aware' MPI library).
\begin{figure}[htbp]
\centering
\noindent\begin{minipage}{.9\textwidth}
\begin{lstlisting}
  call MPI_Comm_split_type (MPI_COMM_WORLD,MPI_COMM_TYPE_SHARED,0,
&                           MPI_INFO_NULL,comm_shared,ierr)
  call MPI_Comm_size (comm_shared, nprocsh, ierr)
  call MPI_Comm_rank (comm_shared, iprocsh, ierr)
  igpu = MODULO(iprocsh, nprocsh)
!$acc set device_num(igpu)
\end{lstlisting}
\end{minipage}
\noindent\begin{minipage}{.9\textwidth}
\begin{lstlisting}
!$acc host_data use_device(y) if_present
  call MPI_Allreduce (MPI_IN_PLACE,y,n,MPI_DOUBLE,MPI_SUM,MPI_COMM_WORLD,ierr)
!$acc end host_data  
\end{lstlisting}
\end{minipage}
\caption{Top:  Example approach to selecting the GPU device to use for a particular MPI rank.  Bottom: Example of using CUDA-aware MPI to avoid CPU-GPU transfers for MPI communications.\label{fig:accexp3}} 
\end{figure}

These are just a few examples of the OpenACC directives needed for the MAS code.  We refer the reader to Ref.~\cite{CAPLAN_DEVBLOG} for many more detailed examples.

\subsection{Challenges}
\label{sec:difficult}
Implementing OpenACC comes with its fair share of challenges.  As a relatively new programming API, there are often delays in clear documentation for new features, delays in compilers implementing the latest features, as well as inevitable compiler bugs.  We encountered all of these during our development of OpenACC into MAS, adding difficulty to the otherwise fairly straight-forward implementation.  For example, in order to work around a few bugs in the PGI compiler, we had to modify more CPU code than otherwise required.  We also tried using very new language features that offered even simpler coding, but in the end, needed to use  standard features for the code to function properly with the compiler.  

In addition to the challenges posed by compiler implementations, we also encountered difficulties running the resulting code on the HPC resources available to us.  For example, certain new features of OpenACC that were required for MAS were not available in the older versions of the compiler installed on the cluster, and updating the compiler took some time.  Also, some MPI libraries were not up-to-date, not allowing some required GPU features.  Although we were able to get everything working eventually through the help of the centers' staff, these problems added non-trivially to the development time.  We note that HPC systems specifically designed for GPU computing (such as the Summit system at ORNL \cite{SUMMIT}), are not likely to exhibit these issues.

\subsection{Summary of code modifications}
The current implementation of OpenACC into MAS for zero-beta runs required modifying less than 5\% of the code.  The total number of added OpenACC comments amounted to less than 1.5\% the length of the original code.  Certain compiler bugs and currently lacking features of OpenACC account for much of the remaining code modifications, along with optional CPU code optimizations implemented during the development (since only the zero-beta features of MAS were accelerated, an implementation of the full feature-set of MAS will necessarily require additional OpenACC directives).  The accelerated code only uses directives, keeping the code fully portable and able to be compiled with any previously used CPU compiler.   

\section{Timing Results}
\label{sec:timings}
Comparing CPU and GPU performance of a given code in a `fair' manner can be difficult since the optimal implementation of the algorithms are often unique for each hardware (either explicitly in the code, or implicitly through the compiler).  In addition, different algorithms may perform better on one hardware than another, (e.g. the PC2 preconditioner of our PCG solvers is more suited for a CPU, while PC1 is better suited for a GPU).  However, our goal is not to benchmark hardware, but rather to test the effective “time-to-solution” for a given problem using GPU acceleration versus the original CPU code.  Therefore, we first test each available algorithm on each hardware configuration using the best suited compiler and select the fastest running algorithm for each.  Implementation of new algorithms could potentially improve performance on the CPU and GPU, but here we want to focus on what can be achieved adding GPU acceleration to our current code base in a portable way.  

To measure the effective “time-to-solution”, we record the absolute wall clock time of our runs including all CPU-code setup, I/O, CPU-GPU transfers, etc. While we do not include the time spent waiting in the scheduling queue of HPC clusters, it should not be overlooked when evaluating the advantage of an 'in-house' solution.  The timings are recorded using calls to the MPI library's timing API routines, and all simulations are computed in double-precision (FP64). 

\subsection{HPC environment}
\label{sec:testcasehpc}
In order to give a broad picture of performance, we run the simulations on multiple generations of hardware.  
For the CPU code, we run on five generations of Intel processors (from Sandy Bridge (rel. 2012) up to Skylake (rel. 2017)) on NASA's NAS Pleiades and Electra systems.  For the GPU code, we run on three generations of NVIDIA GPUs (Kepler (rel. 2014) up to Volta (rel. 2017)) on SDSC's Comet and NVIDIA's PSG machines.  Each node on the CPU systems are dual-socket (2 CPUs per node), while each GPU node has four GPUs per node (the Kepler K80 GPU is a dual-GPU card and therefore each node has two K80 GPU cards to make up the four GPUs).  We also test each code using an in-house desktop PC with an Intel Broadwell CPU and a NVIDIA TitanXP GPU.  Detailed specifications for all CPU and GPU hardware and software configurations (including compiler versions and flags, driver versions, etc.) are given in the Appendix. 

\subsection{Algorithm comparisons}
\label{sec:timealgcomp}
\begin{figure}[htbp]
\centering
\includegraphics[height=4in]{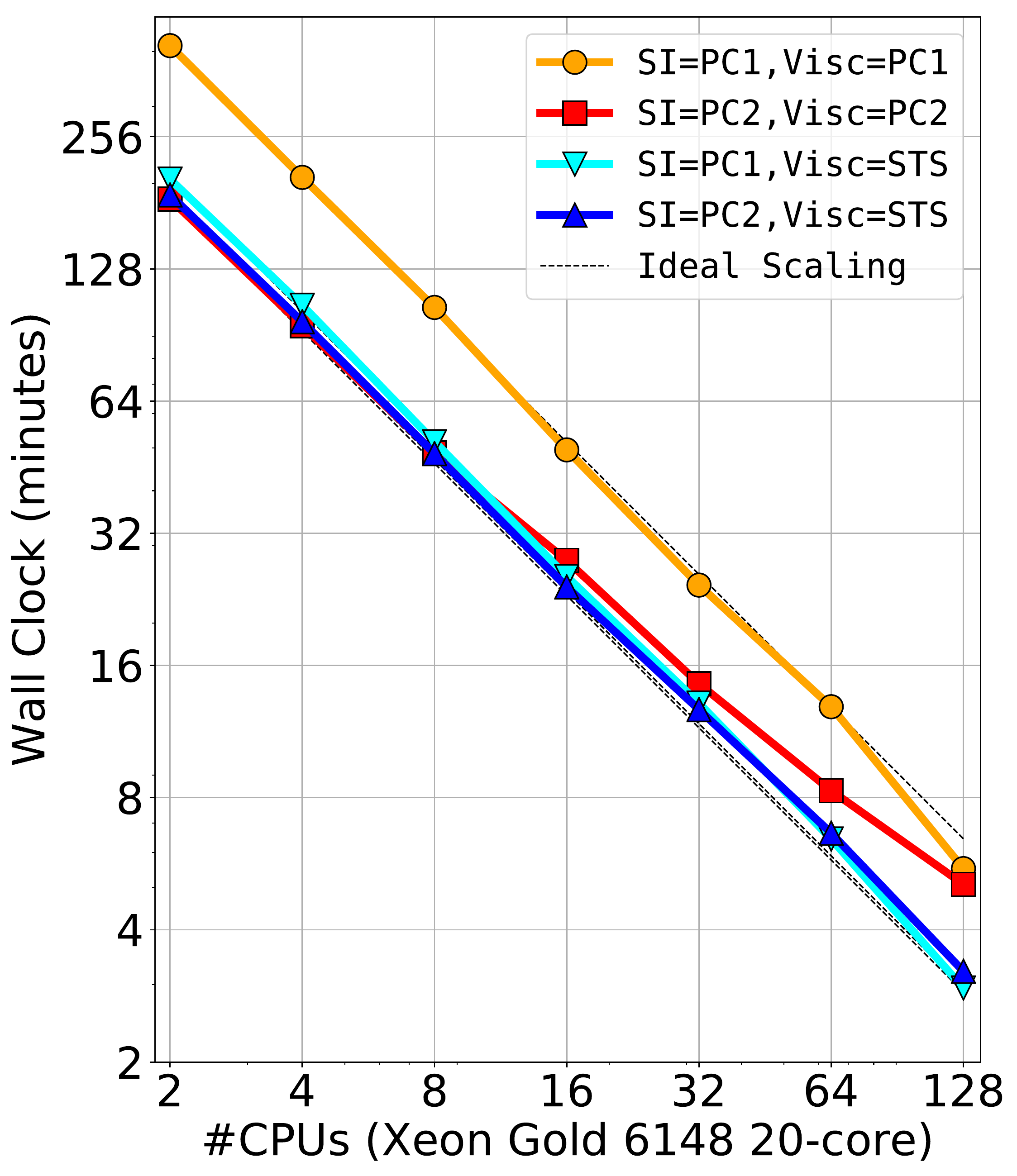}
\includegraphics[height=4in]{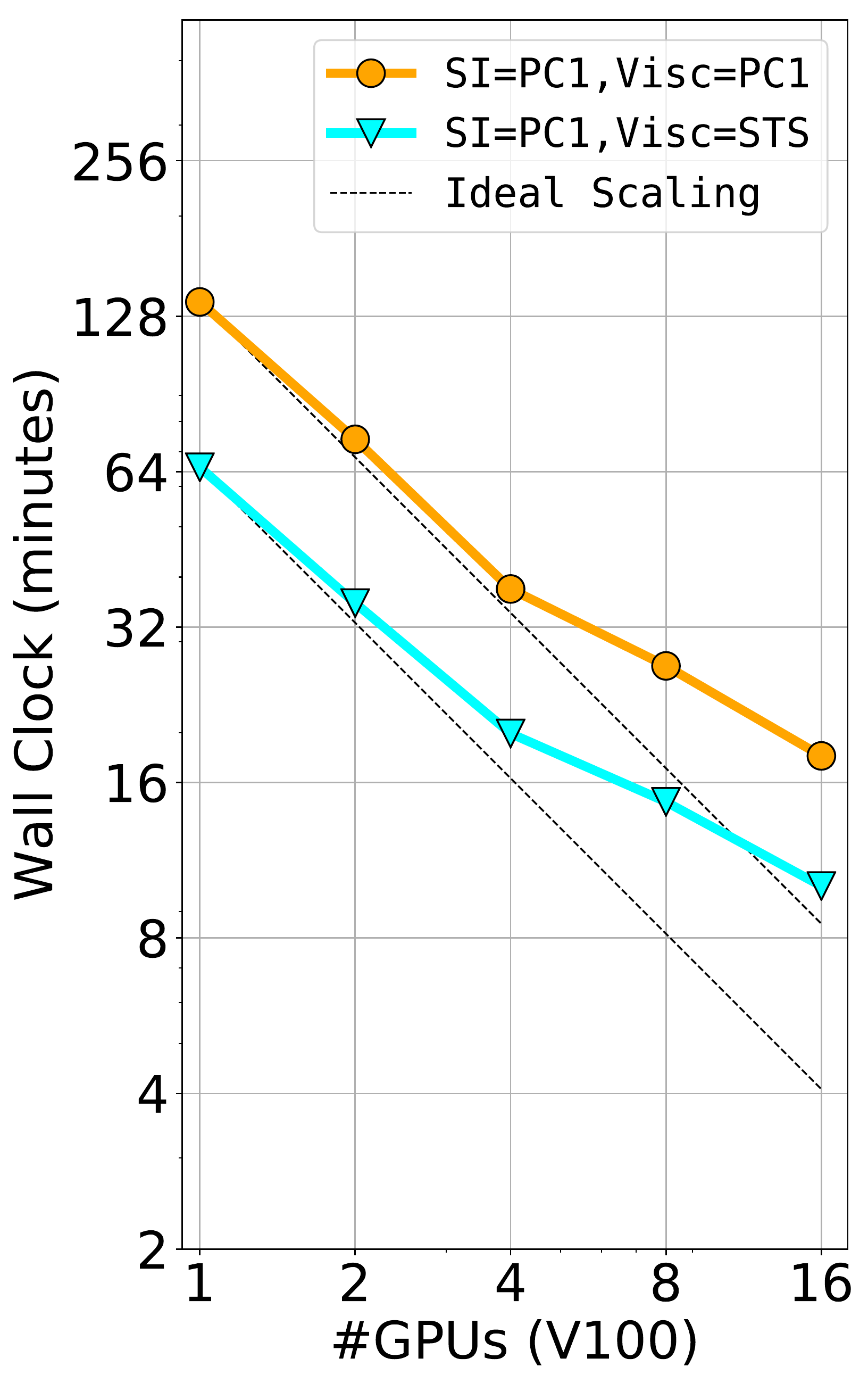}
\caption{Comparison of run times and scaling for the algorithm choices in Sec.~\ref{sec:algs} for the test run of Sec.~\ref{sec:testcase} using CPUs (left) and GPUs (right).  The CPU timings were performed on server nodes with 2 Skylake 20-core CPUs per node, while the GPU timings used servers with 4 V100 GPUs per node (see Appendix for details).\label{fig:timings_alg_compare}} 
\end{figure}

Here, we test each available algorithm option from Sec.~\ref{sec:algs} for the code running on CPUs and GPUs to determine the most efficient for each.  We use the best available model of CPU and GPU for the timings (in this case, the Skylake CPU and Volta GPU) and record the performance and scaling using the test run of Sec.~\ref{sec:testcase}.  The results are shown in Fig.~\ref{fig:timings_alg_compare}.
The CPU results show that using only the PCG solver with PC2 runs almost twice as fast as when using PC1 for low number of CPUs.  However, for the highest numbers of CPUs, the improved scaling of the PC1 preconditioner makes the two solvers run in similar time.  Using STS for viscosity exhibits perfect scaling, and the run time of using PC1 or PC2 for the semi-implicit solves yields similar results.  However, since the PC2+STS does exhibit slightly faster times overall than PC1+STS, we will use PC2+STS for our CPU comparison tests.

Both algorithm choices (PC1+PC1 and PC1+STS) for the GPU show similar scaling, with PC1+STS being about twice as fast as PC1+PC1.  Therefore, we choose the PC1+STS algorithm for our GPU comparison runs.  There is a noticeable reduction in scaling past 4 GPUs (although the runs still reduces in time with more GPUs).  Since each compute node contains 4 GPUs, this reduction appears to be an inter-node scaling problem.  While it is possible that  the test problem is simply too small to scale to many GPUs, we note that the PSG cluster and Comet do not have direct RDMA GDR enabled for our choice of MPI library.  This means that even when using CUDA-aware MPI, all communications require the transfer of data to and from the GPU.  Since the amount of data transferred in the MPI calls is much larger in these runs compared to comparably-timed CPU runs (due to much fewer MPI ranks leading to larger grid halo sizes), the lack of RDMA is likely a contributing factor to the poor scaling of the inter-node runs.

We have chosen two unique algorithms for our CPU and GPU comparison timings.  While this may seem like an unfair comparison, as mentioned, we are only interested in the effective `time-to-solution' we can achieve on both platforms, not in an apples-to-apples comparison of the hardware.  However, since the run times of the PC2+STS and PC1+STS algorithms on the CPUs are quite similar, the CPU-GPU comparisons do happen to be close to a direct comparison in this case.

\subsection{Hardware comparisons}
\label{sec:timehw}
\begin{figure}[tbp]
\centering
\includegraphics[height=5in]{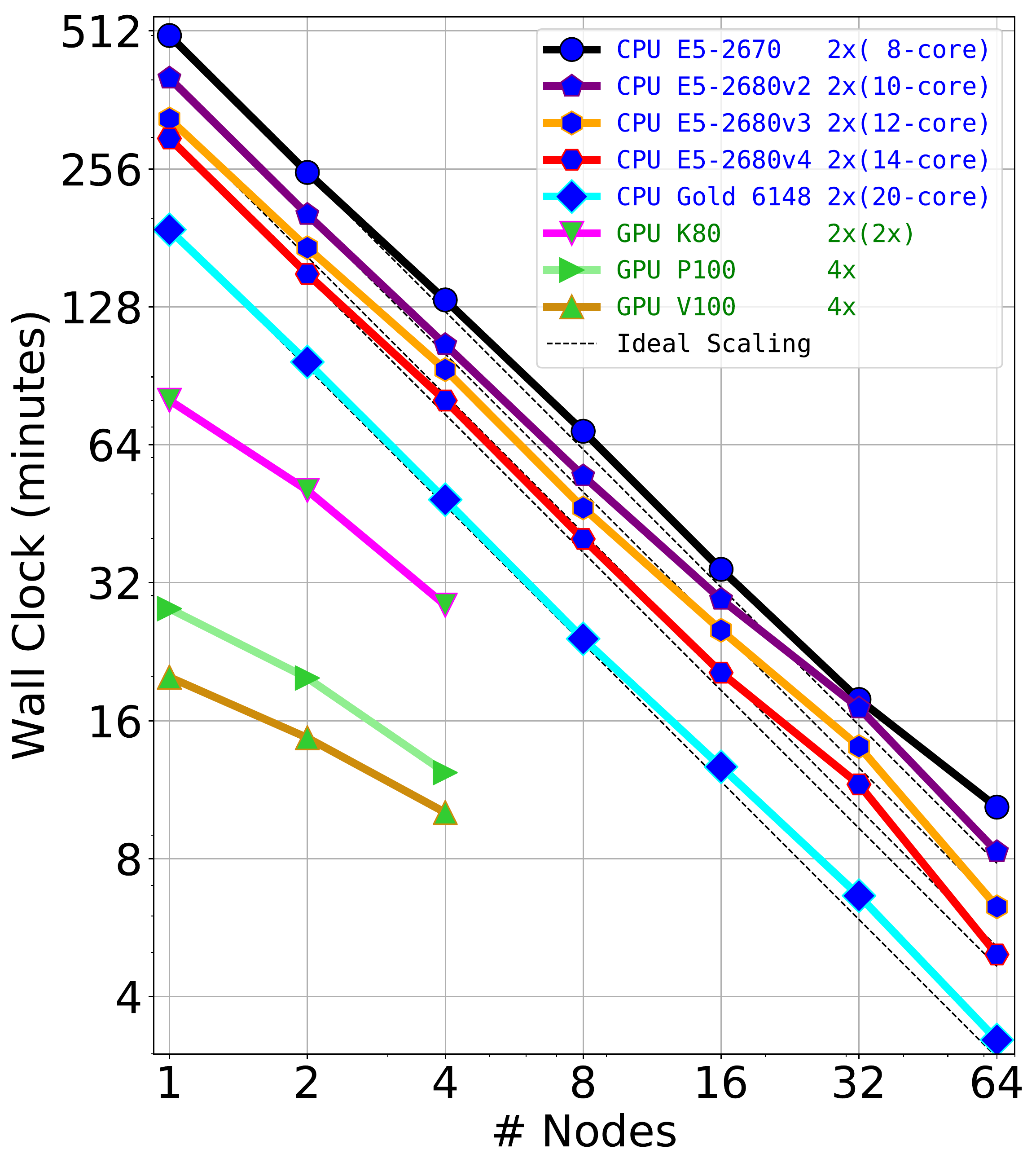}
\caption{Comparison of wall clock times for the test run using various generations of CPU and GPU hardware.  The CPU runs use the PC2+STS algorithm, while the GPU runs use the PC1+STS algorithm (see Sec.~\ref{sec:timealgcomp}).  Details of the hardware and software environment used is given in the Appendix.\label{fig:timings_hw_compare}} 
\end{figure}

Using the best chosen algorithms for the CPU and GPU, we compare wall clock times for several generations of CPU and GPU architectures.  Since a single GPU server could be configured with up to 16 GPUs and a single CPU server could be configured with up to 4 CPU multi-core processors, it is difficult to directly compare timing results.  In line with our practical approach, we choose to compare timing results as a function of the number of compute nodes used on the HPC systems (for CPUs, dual-socket configurations (2 CPU processors per node) and for GPUs, 4 GPUs per node).  Figure~\ref{fig:timings_hw_compare} shows the results for 5 models of CPU and 3 models of GPU.
The CPU runs all scale nearly perfectly as expected from the tests in Sec.~\ref{sec:timealgcomp}, while all three GPU models show similar scaling problems between nodes.  However, even with the reduced scaling, for a given number of compute nodes, the GPU runs are much faster than those of the CPUs.  Allocation limits prevented us from testing past 4 GPU-enabled nodes.  Using just 1 node of 4xV100s yields a run time that would require more than 8 40-core SkyLake CPU nodes (or nearly 32 SandyBridge nodes) to achieve.  Thus, the GPU nodes exhibit excellent compact performance.

\subsection{Single System results}
\label{sec:timess}
\begin{figure}[htbp]
\centering
\includegraphics[width=\textwidth]{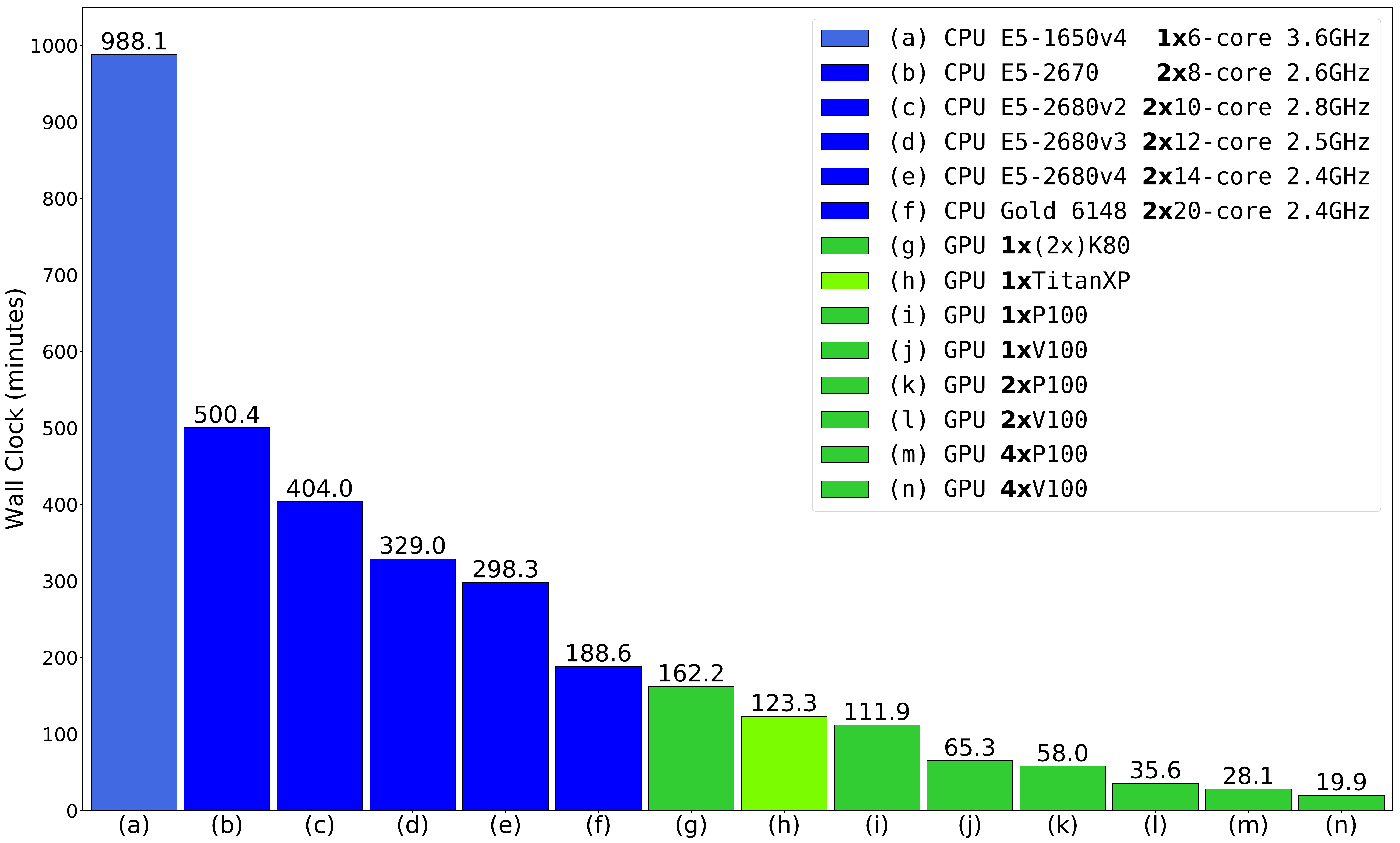}
\caption{Single system wall clock comparisons for the test run on various CPU and GPU hardware.  The advantages of the GPU implementation is clearly seen.  Although we only had access to single GPU systems up to 4 GPUs, we note that up to 16xGPU single servers are available.\label{fig:timings_single_server}} 
\end{figure}

One of the goals for adding GPU acceleration to MAS is to allow small-to-medium simulations to run `in-house' that otherwise would require the use of an HPC CPU cluster.  To illustrate this, in Figure~\ref{fig:timings_single_server} we show timing results for CPUs and GPUs on a single server (limited to those available to us).  We also add CPU and GPU timings for a consumer-level desktop PC equipped with a consumer-level NVIDIA GPU for comparison.  All hardware and software details are listed in the Appendix.  
We see that while a dual-socket Skylake server runs the test problem over 5 times faster than our local desktop PC, a single server of 4xV100 GPUs runs the same test almost 50 times faster (9.5x faster than the Skylake node).  When considering that one is able to configure a single server with up to 16 GPUs, the advantages of the GPU implementation are compelling.  We also see that the consumer-level GPU (TitanXP) performs almost as well as the server counterpart of the same architecture (P100).  This may seem surprising at first, since, while the TitanXP has a clock rate $\sim$15\% faster than the P100, it contains 15x less double-precision compute cores.  We can explain the result by noting that the  memory bandwidth of the two GPUs are within 25\% of each other.  Since the algorithms in MAS are all `memory-bound' (and never reach close to the theoretical maximum FLOP/s on any architecture), it is not too surprising that the performance of the two cards are within 10\% of each other.  This performance similarity opens the door for very inexpensive GPU computing with the MAS code (or other codes with memory-bound algorithms), greatly increasing the relevance of the single-server results.
\section{Conclusions}
\label{sec:conclusion}
We have successfully added GPU acceleration to the MAS code running in zero-beta mode in a fully portable manner using OpenACC.  The use of OpenACC allows us to maintain a single source code that can be compiled to either multi-CPU or multi-GPU platforms, with minimal modifications to the original code.  This avoids the steep learning curves and development interference many associate with GPU programming.

Using the best algorithm choice for each hardware, we showed that one can achieve performance using a few GPUs on one server that requires numerous CPU servers to match.  This opens the door for running small-to-medium resolution simulations `in-house', greatly assisting development and research.  The OpenACC implementation will also allow the MAS code to utilize the latest state-of-the-art HPC multi-GPU systems (e.g. the Summit machine at ORNL \cite{SUMMIT}).  

Efforts are underway to continue the OpenACC implementation, with the goal of eventually supporting MAS's full set of features for use with a wide range of solar physics and space weather applications.
\section*{Acknowledgments}
Work at Predictive Science was supported by AFOSR, NASA, and NSF.  Computational resources provided by the NSF-supported XSEDE program for use of San Diego Supercomputer Center's COMET system, NASA NAS for use of the Pleiades and Electra systems, and NVIDIA Corporation for use of their PSG test system.

\section*{References}
\bibliographystyle{iopart-num}
\bibliography{Caplan_ASTRONUM18}
\newpage
\section*{Appendix: Hardware and software details}
\label{sec:app}
The hardware and software details used for the CPU tests are given in Table~\ref{tab:sysinfoCPU} and those for the GPU tests in Table~\ref{tab:sysinfoGPU}.  The compiler flags used for each test are shown in Table~\ref{tab:compflags}.
\begin{table}[htbp]
\centering
\resizebox{\textwidth}{!}{%
\begin{tabular}{c|c|c|c|c|c|c|}
\cline{2-7}
& \multicolumn{5}{|c|}{NASA NAS Pleiades \& Electra} & Local Desktop\\
\hline
\multicolumn{1}{|c|}{Compiler} & \multicolumn{5}{|c|}{Intel 2018.3.222} & GNU 5.4.0\\
\hline
\multicolumn{1}{|c|}{MPI Library} & \multicolumn{5}{|c|}{SGI MPT 2.15r20} & OpenMPI 2.1.1\\
\hline
\multicolumn{1}{|c|}{CPU Family} & Sandy Bridge & Ivy Bridge & Haswell & Broadwell & Skylake & Broadwell\\
\hline
\multicolumn{1}{|c|}{Instruction Set} & \multicolumn{2}{|c|}{AVX} & \multicolumn{2}{|c|}{AVX2} & {AVX512} & AVX2\\
\hline
\multicolumn{1}{|c|}{CPU Model} & E5-2670 & E5-2680v2 & E5-2680v3 & E5-2680v4 & Gold 6148 & E5-1650v4\\
\hline
\multicolumn{1}{|c|}{CPU Clock Rate} & 2.6 GHz & 2.8 GHz & 2.5 GHz & 2.4 GHz & 2.4 GHz & 3.6 GHz\\
\hline
\multicolumn{1}{|c|}{\#Sockets x \#Cores} & 2x8 & 2x10 & 2x12 & 2x14 & 2x20 & 1x6\\
\hline
\multicolumn{1}{|c|}{Memory Bandwidth} & 51.2 GB/s & 59.7 GB/s & 68 GB/s & 76.8 GB/s & 128 GB/s & 76.8 GB/s\\
\hline
\end{tabular}}
\caption{System hardware and software configuration details for CPU results.\label{tab:sysinfoCPU}}
\end{table}

\begin{table}[htbp]
\centering
\resizebox{0.8\textwidth}{!}{%
\begin{tabular}{c|c|c|c|c|}
\cline{2-5}
& \multicolumn{2}{|c|}{SDSC COMET} & NVIDIA PSG & Local Desktop\\
\hline
\multicolumn{1}{|c|}{Compiler} & \multicolumn{2}{|c|}{PGI 18.4} & PGI 18.3 & PGI 18.4\\
\hline
\multicolumn{1}{|c|}{MPI Library} & \multicolumn{2}{|c|}{OpenMPI 2.1.2} & OpenMPI 1.10.7 & OpenMPI 2.1.2\\
\hline
\multicolumn{1}{|c|}{CUDA Version} & \multicolumn{4}{|c|}{CUDA 9.1}\\
\hline
\multicolumn{1}{|c|}{Driver Version} & \multicolumn{2}{|c|}{367.48} & 396.26 & 410.48\\
\hline
\multicolumn{1}{|c|}{GPU Family} & Kepler & Pascal & Volta & Pascal\\
\hline
\multicolumn{1}{|c|}{GPU Model} & K80 & P100 & V100 & TitanXP\\
\hline
\multicolumn{1}{|c|}{\#GPUs/node} & 2x2 & 4 & 4 & 1\\
\hline
\multicolumn{1}{|c|}{GPU Clock Rate} & 0.56 GHz & 1.33 GHz & 1.38 GHz & 1.58 GHz\\
\hline
\multicolumn{1}{|c|}{\#FP64 Cores/GPU} & 2x832 & 1792 & 2560 & 120\\
\hline
\multicolumn{1}{|c|}{Memory Bandwidth/GPU} & 2x240 GB/s & 732 GB/s & 900 GB/s & 547.6 GB/s\\
\hline
\end{tabular}}
\caption{System hardware and software configuration details for GPU results.\label{tab:sysinfoGPU}}
\end{table}

\begin{table}[htbp]
\centering
\resizebox{0.7\textwidth}{!}{%
\begin{tabular}{|l|l|}
\hline
Compiler & Flags
\\
\hline
\hline
Intel 2018.3.222 (CPU) & {\tt -O3 -heap-arrays -fp-model precise} \\
& {\tt -axCORE\_AVX512 -xSSE4.2 -DNDEBUG}
\\
\hline
GNU 5.4.0 (CPU) & {\tt -O3 -mtune=native -DNDEBUG}
\\
\hline
PGI 18.3 \& 18.4 (GPU) & {\tt -O3 -ta=tesla:cuda9.1,cc60,cc70 -DNDEBUG}
\\
\hline
\end{tabular}}
\caption{Compiler flags used to in compiling MAS for the test runs.\label{tab:compflags}}
\end{table}

\end{document}